\documentclass[traditabstract]{aa}
\usepackage{txfonts}
\usepackage{amssymb}
\usepackage{graphicx}
\begin{document}
\title{Constructing the secular architecture of the solar system II: The terrestrial planets}
\author{R. Brasser \inst{1}
\and A. Morbidelli \inst{1}
\and R. Gomes \inst{2}
\and K. Tsiganis \inst{3}
\and H. F. Levison \inst{4}}
\institute{Dep. Cassiopee, University of Nice - Sophia Antipolis, CNRS, Observatoire de la C\^{o}te d'Azur; Nice, France
\and Observat\'{o}rio Nacional; Rio de Janeiro, RJ, Brasil
\and Department of Physics, Aristotle University of Thessaloniki; Thessaloniki, Greece
\and Southwest Research Institute; Boulder, CO, USA}
\date{submitted: 13 July 2009; accepted: 31 August 2009}
\abstract{We investigate the dynamical evolution of the terrestrial planets during the planetesimal-driven migration of the giant planets. A basic assumption of this work is that giant planet migration occurred after the completion of terrestrial planet formation, such as in the models that link the former to the origin of the Late Heavy Bombardment. The divergent migration of Jupiter and Saturn causes the $g_5$ eigenfrequency to cross resonances of the form $g_5=g_k$ with $k$ ranging from 1 to 4. Consequently these secular resonances cause large-amplitude responses in the eccentricities of the terrestrial planets if the amplitude of the $g_5$ mode in Jupiter is of the order of the current one. We show that the resonances $g_5=g_4$ and $g_5=g_3$ do not pose a problem if Jupiter and Saturn have a fast approach and departure from their mutual 2:1 mean motion resonance. On the other hand, the resonance crossings $g_5=g_2$ and $g_5=g_1$ are more of a concern as they tend to yield a terrestrial system incompatible with the current one, with amplitudes of the $g_1$ and $g_2$ modes that are too large. We offer two solutions to this problem. The first uses the fact that a secular resonance crossing can also damp the amplitude of a Fourier mode if the latter is large originally. We show that the probability of the $g_5=g_2$ resonance damping a primordially excited $g_2$ mode in the Earth and Venus is approximately 8\%. Using the same mechanism to additionally ensure that the $g_5=g_1$ resonance keeps the amplitude of the $g_1$ mode in Mercury within 0.4 reduces the overall probability to approximately 5\%. These numbers, however, may change for different initial excitations and migration speed of the giant plants. A second scenario involves a 'jumping Jupiter' in which encounters between an ice giant and Jupiter, without ejection of the former, cause the latter to migrate away from Saturn much faster than if migration is driven solely by encounters with planetesimals. In this case, the $g_5=g_2$ and $g_5=g_1$ resonances can be jumped over, or occur very briefly. We show that in this case the terrestrial system can have dynamical properties comparable to what is exhibited today. In the framework of the Nice model, we estimate that the probability that Jupiter had this kind of evolution to be approximately $6\%$.}
\titlerunning{Secular architecture of the terrestrial planets}
\maketitle

\section{Introduction}
In this paper we continue our effort to understand the origin of the orbital architecture of the planets of the solar system. In a previous work (Morbidelli {{\textit{et al}}}., 2009), which is henceforth called Paper I, we analysed the secular architecture of the outer solar system and concluded that, in addition to radial migration, encounters between Saturn and one of the ice giants needed to have occurred in order to explain the current properties of the secular dynamics of the outer solar system. Here we investigate the orbital evolution of the terrestrial planets during the changes that occurred in the outer Solar System. \\

A prerequisite for this study is a discussion of whether or not the terrestrial planets existed at the time the giant planets changed
their orbital architecture. According to our best models (Chambers \& Wetherill, 1998; Agnor {{\textit{et al}}}., 1999; Chambers, 2001; Raymond {{\textit{et al}}}., 2004, 2005, 2006, 2007; O'Brien {{\textit{et al}}}., 2006, Kenyon \& Bromley, 2006) the terrestrial
planets formed by collisions among a population of intermediate objects called planetary embryos, with masses of the order of 0.01--0.1 Earth masses. This process should have lasted several tens of millions to a hundred million years, in agreement with modern results from the analysis of radioactive chronometers (Touboul {{\textit{et al}}}., 2007; All\`egre {{\textit{et al}}}., 2008). In contrast, the giant planets had to have formed in a {{\textit{few}}} million years, otherwise they could not have trapped the gas from the primordial solar nebula, which typically disappears on this timescale (e.g. Haisch {{\textit{et al}}}., 2001). Once triggered, the migration and the other changes in the orbital structure of the giant planets typically take a few tens of million years (Gomes {{\textit{et al}}}., 2004). So, putting all these timescales together, it is legitimate to think that, by the time the terrestrial planets completed their formation, the giant planets were already on their current orbits. If this is indeed the case, then there is no object for the present study. \\

However, there is an emerging view that the re-organisation of the orbital structure of the giant planets might have had a delayed start (Gomes {{\textit{et al}}}., 2005; Strom {{\textit{et al}}}., 2005; see also Levison {{\textit{et al}}}., 2001) and therefore it might have postdated the formation of the terrestrial planets. The reason to think so is that the terrestrial planets underwent a Late Heavy Bombardment (LHB) of small bodies. The temporal evolution of this bombardment is still subject of debate (see for instance Hartman {{\textit{et al}}}., 2000, for a review), but the majority of the evidence points to a cataclysmic nature of the LHB, i.e. to a spike in the cratering rate occurring approximately 3.85~Gyr ago (e.g. Ryder {{\textit{et al}}}., 2000) i.e. approximately 600~Myr after terrestrial planet formation. If this is true, then something ``major'' had to happen in the solar system at that time, and the late change in the orbital structure of the giant planets seems to be a plausible explanation (although, see Chambers, 2007, for an alternative scenario that does not involve a change in the giant planets' orbits). \\

The study of the evolution of the terrestrial planets during the putative changes of the giant planets' orbits is therefore a key to
unveiling the real evolution of the Solar System. For instance, if we find that the current orbital architecture of the terrestrial planets is incompatible with giant planet migration, then explanations of the LHB based on a late migration of Jupiter and Saturn (Gomes {{\textit{et al}}}. 2005; Strom {{\textit{et al}}}, 2005) should be rejected. If, on the contrary, we find that some giant planet evolutions, consistent with the constraints of Paper~I, are also consistent with the current architecture of the terrestrial planets, then we have made another important step on the way of building a coherent and consistent scenario of the dynamical history of the Solar System. This is precisely the purpose of the present paper. \\

In the next section, we explain what are the challenges set by the migration and the orbital excitation of the giant planets on the
stability of the terrestrial planets. Then, in section~3 we look in detail on the evolution of the Earth-Venus couple and of Mercury and how their current secular dynamics could be achieved or preserved. Section~4 is devoted to Mars. Section~5 will briefly discuss the evolution of the inclinations. Finally, Section~6 is devoted to the discussion on the relative timing of  giant planets 
migration versus terrestrial planets formation. The conclusions at the light of our result are presented there as well.

\section{Giant planet migration and the evolution of the terrestrial planets}
\subsection{Overview}
The evolution of the eccentricities and longitudes of perihelia of the four terrestrial planets can be described, in first approximation, by the Lagrange--Laplace solution of the secular equations of motion (see Chapter 7 in Murray \& Dermott, 1999):
\begin{eqnarray}
e_i\cos\varpi_i&=&\sum_k M_{i,k}\cos\alpha_k\cr 
e_i\sin\varpi_i&=&\sum_k M_{i,k}\sin\alpha_k, 
\label{Lagrange}
\end{eqnarray}
where the index $i$ refers to the planet in consideration and $k$ ranges from 1 to 8 with $\alpha_k=g_kt+\beta_k$. The $g_k$ are called proper frequencies of the secular perihelion-eccentricity motion of the planets. The frequencies $g_5$ to $g_8$ are those characterising the system of the giant planets Jupiter to Neptune; they appear also in the equations describing the evolution of the terrestrial planets because the latter are perturbed by the former. In reality, the amplitudes of the terms corresponding to $g_6$, $g_7$ and $g_8$ are very small in the terrestrial planets and can be neglected to first approximation. However this is not the case for the terms corresponding to $g_5$. The frequencies $g_1$ to $g_4$ are proper of the terrestrial planets. Table~\ref{sjs} reports the current values of $g_1$ to $g_5$ and Table~\ref{svemnow} gives the coefficients $M_{i,j}$ with non-negligible amplitude in equation (\ref{Lagrange}). The coefficients of the terms with frequencies $g_6$ to $g_8$ are omitted because their amplitudes are much smaller than the ones given here. Data for both tables are taken from Laskar (1990) and Morbidelli (2002). As one sees from Table~\ref{sjs}, the frequencies $g_1$ to $g_4$ can be partitioned into two groups: $g_1$ and $g_2$ are small, of the order of 5--7 $\arcsec$/yr$^{-1}$, but nevertheless they are larger than the current value of $g_5$; on the other hand, $g_3$ and $g_4$ are much larger, of the order of 17--18 $\arcsec$/yr$^{-1}$.\\

Notice that, because the terrestrial planets exhibit weakly chaotic dynamics (Laskar, 1990), the frequencies $g_1$ to $g_4$ and their amplitudes and corresponding phases are not constant with time. The changes are particularly relevant for the frequency $g_1$. Thus, Tables~\ref{sjs} and~\ref{svemnow} should be considered as indicative, and only reflect the current dynamics. Their values might have been different
in the past, even since the giant planets achieved their current orbital configuration. Consequently, the maximal eccentricities that the planets attain during their secular oscillation could change over time. For instance, the current maximal eccentricity of Venus is 0.072 (as it can be seen by adding together the absolute values of the coefficients of the second line in Table~\ref{svemnow}). However, over the last 4~Gyr, the eccentricity of Venus had a 10\% probability to exceed 0.09; similarly, with the same probability the eccentricity of Mercury could have exceeded 0.4, that of the Earth 0.08 and that of Mars 0.17 (Laskar, 2008). Correia \& Laskar (2004) argued that some time in the past the eccentricity of Mercury had to be larger than 0.325 to allow it to be captured in its 3:2 spin-orbit resonance. As Mercury should have been in synchronous rotation before the formation of the Caloris basin (Wieczorek {{\textit{et al}}}., 2009), one of the latest big impact events recorded on Mercury, this high-eccentricity phase should have occurred after the LHB.  \\

\begin{table}
\begin{tabular}{ccc}
 Frequency & Value ($\arcsec$/yr) & $\beta$ ($\degr$)\\ 
\\\hline\\
$g_1$ &  5.60 & 112.08 \\
$g_2$ &  7.46 & 200.51 \\
$g_3$ & 17.35 & 305.12 \\
$g_4$ & 17.92 & 335.38 \\
$g_5$ &  4.26 &  30.65
 \\\hline \\
\end{tabular}
\caption{Frequencies and phases for Mercury to Jupiter on their
  current orbits.
}
\label{sjs}
\end{table}
\begin{table}
\begin{tabular}{c|ccccc}
$_j\setminus^k$\quad\hbox{} & 1& 2& 3 & 4 & 5 \\
\\\hline\\
1 & 0.1854 & -0.0277 & 0.0015 & -0.00143 & 0.0363 \\
2 & 0.0067 & 0.0207& -0.0117 & 0.0135 & 0.0196\\ 
3 & 0.0042 & 0.0161 & 0.00941 & -0.0132 & 0.0189 \\  
4 & 0.0007 & 0.00291 & 0.0401 & 0.0490 & 0.0203\\\hline \\
\end{tabular}
\caption{Coefficients $M_{j,k}$ of the Lagrange--Laplace solution for
  the terrestrial planets. 
}
\label{svemnow}
\end{table}

\subsection{Evolution of $g_5$ and its implications}
\begin{figure}
\resizebox{\hsize}{!}{\includegraphics[angle=-90]{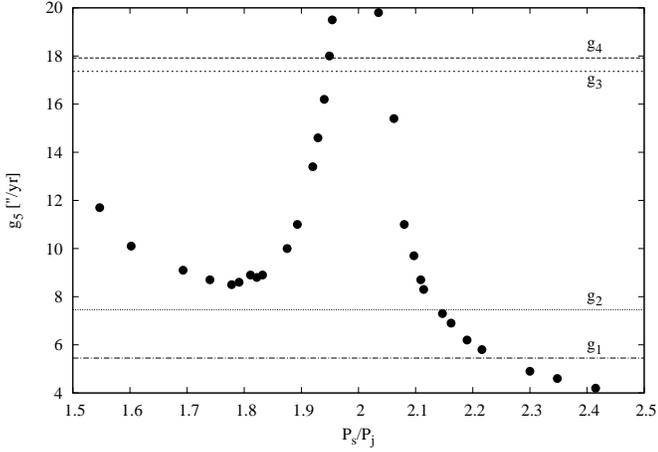}}
\caption{The frequency $g_5$ as a function of $P_S/P_J$. The
horizontal lines denote, from top to bottom, the values of $g_4$,
$g_3$, $g_2$ and $g_1$.}
\label{g5} 
\end{figure}

When Jupiter was closer to Saturn, the value of the $g_5$ frequency had to be higher. Fig.~\ref{g5} shows the value of $g_5$ as a function of the orbital period ratio between Saturn and Jupiter ($P_S/P_J$). The values of $g_5$ have been obtained by numerical Fourier analysis of the outputs of a sequence of 1 Myr integrations of the Jupiter-Saturn pair. The two planets were migrated from 
just outside their 3:2 mean motion resonance to a final, pre-determined period ratio, and subsequently the end result of that migration run was integrated for 1 Myr to obtain the Fourier spectrum. \\

If the period ratio between Saturn and Jupiter had evolved from $P_S/P_J\lesssim 2.15$ to its current value, at least the secular resonances $g_5=g_2$ and $g_5=g_1$ had to be crossed at some time, as one can see from the figure. This effect has already been pointed out by Agnor (2005) and Agnor \& Lin (2007). For reference, the minimal amplitude of planet migration deduced by Malhotra (1995) from the analysis of the Kuiper belt sets the initial $P_S/P_J$ equal to $\sim 2.05$; Minton \& Malhotra (2009) in their recent analysis of the evolution of the asteroid belt, also adopted this initial orbital period ratio. Figure~\ref{g5} shows that the $g_5$ frequency becomes high if Jupiter and Saturn are very close to their mutual 2:1 resonance ($P_S/P_J=2$). This effect is well known and is due to the divergence of the quadratic terms in the masses of the two planets, which are generated when the equations of motion are averaged over the orbital periods (see for instance Kne\v{z}evi\'{c} {{\textit{et al}}}., 1991). As a consequence, if Jupiter passed through or was originally close to the 2:1 resonance with Saturn, as in the Nice model (Tsiganis {{\textit{et al}}}., 2005; Gomes {{\textit{et al}}}., 2005; Morbidelli {{\textit{et al}}}., 2007) or in the scenario of Thommes {{\textit{et al}}}. (2007), also the secular resonances $g_5=g_4$ and $g_5=g_3$ had to be crossed. \\

A secular resonance crossing can significantly modify the amplitudes of the proper modes in the Lagrange--Laplace solution, i.e. the $M_{i,k}$'s. In fact, the Lagrange--Laplace solution of the secular dynamics is only a good approximation of the motion when the planets are far away from secular or mean motion resonances. Thus, if the system passes through a resonance $g_5=g_k$, the terrestrial planets follow a Lagrange--Laplace solution before the resonance crossing and another Lagrange--Laplace solution after the resonance crossing, with the two solutions differing mostly in the amplitudes $M_{i,k}$ of the terms with frequency $g_k$. Thus, the crucial question is: are the current amplitudes of the $g_1$ to $g_4$ terms in the terrestrial planets (i.e. the $M$ coefficients in Table~\ref{svemnow}) compatible with secular resonance crossings having occurred? \\

\begin{figure}
\resizebox{\hsize}{!}{\includegraphics[angle=-90]{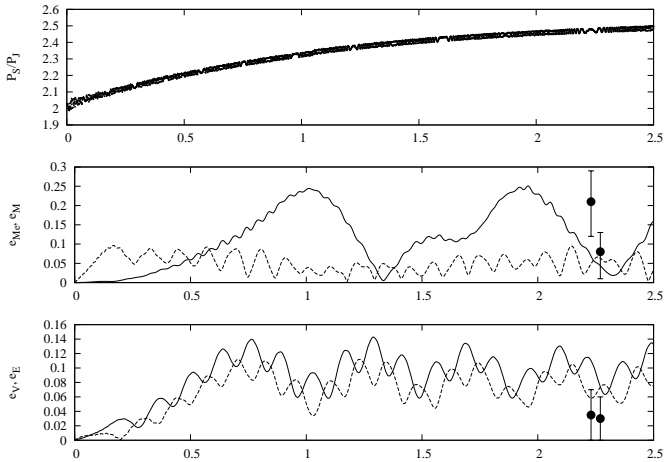}}
\caption{Evolution of the terrestrial planets during the $g_5=g_2$ 
and $g_5=g_1$ resonance crossings. The top panel shows the evolution
of $P_S/P_J$. The middle
panel shows the evolution of the eccentricities of Mercury (solid) and Mars (dashed). The bottom panel
shows the
eccentricities of Venus (solid) and Earth (dashed). The solid circles, accompanied by vertical bars, 
represent the range of variation of the respective planetary element, as given by a 10~Myr simulation 
of the current solar system (data taken from Laskar, 1988). } 
\label{terrg5n} 
\end{figure}

As a demonstration of the effect of secular resonances sweeping through the terrestrial planets system during the migration of Jupiter
and Saturn, we performed a simple experiment: Mercury, Venus, Earth and Mars were placed on orbits with their current semi-major axes and inclinations, but with initial eccentricities equal to zero. Jupiter and Saturn were forced to migrate smoothly from $P_S/P_J\sim 2.03$ to their current orbits (Fig.~\ref{terrg5n}, top), so that $g_5$ sweeps through the $g_4$--$g_1$ range (see Fig.~\ref{g5}). Migration in enacted using the technique discussed in Paper~I, with a characteristic e-folding timescale $\tau=1$ Myr, which is somewhat faster than the fastest time found in Tsiganis {{\textit{et al.}}} (2005). The initial eccentricities and longitudes of perihelion of the giant planets were chosen so that the amplitude of the $g_5$ term in Jupiter was close to that currently observed.  We refer the reader to Paper~I why this is a valid choice. \\

Unlike in Paper~I, in all the simulations presented in this work, we included in the equations of motion the terms resulting from the additional potential
\begin{equation}
V_{\rm GR}=-3\Bigl(\frac{{{\mathcal{G}}}M_{\odot}}{c}\Bigr)^2\frac{a(1-e^2)}{r^3},
\label {GR}
\end{equation}
where $M_\odot$ is the mass of the Sun, ${\cal G}$ is the gravitational constant, $c$ is the speed of light and $r$ is the heliocentric distance. This was done to mimic the effect of General Relativity. Indeed, averaging the potential over the mean anomaly and computing the change in the longitude of pericentre yields

\begin{equation}
 \langle \dot{\varpi}_{GR} \rangle = 3\frac{{{\mathcal{G}}}M_{\odot}}{c^2}\frac{n}{a(1-e^2)} = 0.0383
\Bigl(\frac{1\,{{\rm{AU}}}}{a}\Bigr)^{5/2} \frac{1}{1-e^2}\,\arcsec/yr,
\end{equation}
in accordance with Nobili \& Will (1986). This potential term yields a precession rate of the longitude of perihelion of Mercury of 0.43 $\arcsec$/yr. A more complex post-Newtonian treatment of General Relativity is possible (see for instance Saha \& Tremaine, 1994), but the additional terms account for short periodic effects or secular changes in the mean motions of the planets, so they are not important in our case. Conversely in the current solar system, accounting for equation (\ref{GR}) is important because it increases $g_1$ so that it is further away from the current value of $g_5$ (and hence from the asymptotic value of $g_5$ in a migration simulation like ours), which helps in stabilising the motion of Mercury (Laskar, 2008). \\

Returning to Figure \ref{terrg5n} we see that Mercury's eccentricity reaches 0.25 (solid line; middle panel), which is consistent with its current orbit. The current mean value and range in eccentricity of Mercury is displayed by the first, higher bullet with error bars. We should stress that some other simulations, with a similar set up than this one, led to an eccentricity of Mercury exceeding 0.5. Mars' eccentricity (dashed line; middle panel) is excited up to 0.1 very early in the simulation and then oscillates in the 0-0.1 range i.e. slightly less than in reality (depicted by the second, lower bullet with error bars). Alternatively, Venus acquires a mean eccentricity around 0.1, with a maximal value as large as 0.14 (bottom panel; solid line), while the maximal eccentricity of the Earth exceeds 0.1 (bottom panel; dashed line). Thus, the Earth and Venus become significantly more eccentric than they are, or can be, in the current solar system (Laskar, 2008). \\

The reason for this behaviour is that the amplitudes corresponding to the $g_1$ and $g_2$ frequencies have been strongly excited in Mercury, Venus and Earth by the passage through a resonance with the $g_5$ frequency. Similarly, the $g_4$ mode in Mars is excited very early on by the same mechanism. A Fourier analysis done on a 4~Myr continuation of the simulation, with Jupiter and Saturn on non-migrating orbits, reveals that the amplitude of the $g_2$ term in Venus is about 0.1, i.e. five times larger than the current value. The amplitude of the $g_1$ term in Mercury is also $\sim 0.1$, which is less than its current value. The evolution of Mercury, though, is not only influenced very strongly by Jupiter, but also by Venus.  Thus, the excitation of the $g_1$ term depends sensitively on the phase that the $g_2$ term has in Mercury when the $g_5=g_1$ resonance occurs. Consequently, it is very easy to excite the amplitude of the $g_1$ mode in Mercury to a much larger value by slightly changing the set-up of the simulation so that the time at which the resonance occurs is somewhat different. In addition, if the terrestrial planets are started on co-planar orbits, $g_1$ is faster and therefore closer to $g_2$ and consequently the $g_1$ term is excited more easily because of a quasi-resonance between $g_1$ and $g_2$. Conversely, since $g_5$ passes very quickly through the values of $g_4$ and $g_3$, the amplitudes of the corresponding terms in Mars and the Earth are only moderately excited ($0.015$ for $g_3$ and 0.04 for $g_4$ in Mars). Since the final amplitude of the $g_2$ term is much larger than that of the $g_3$ term, the Earth and Venus are in apsidal libration around $0^\circ$, as explained in Paper~I. \\

Thus, this simulation shows that, at least in the case of a fast migration (see Section~4), the $g_5=g_4$ and $g_5=g_3$ resonances are
not a severe problem. But the $g_5=g_2$ and $g_5=g_1$ resonances, occurring towards the end of the migration, are a more serious concern for reconstructing the current secular architecture of the terrestrial planets. For a fast migration speed the excitation of the $g_2$  mode is a linear function of the migration timescale, $\tau$\footnote{We have verified this in our simulations}. Consequently, given that with $\tau=1$~Myr the amplitude of the $g_2$ term is five times larger than the current value, achieving the 
current excitation would require $\tau = 0.2$ Myr. This value of $\tau$ is unrealistic for planetesimal-driven migration of the giant planets. For example, in the preferred case of Hahn \& Malhotra (1999), in which planet migration is driven by a 50 $M_{\oplus}$ planetesimal disc, it takes more than 30 Myr for the planets to reach their current orbits. Assuming an exponential fit to the migration and allowing the migration to be essentially finished after 3 e-folding times, will yield $\tau \sim 10$ Myr. A similar result can be found in Gomes {{\textit{et al.}}} (2004). The Nice model is the scenario in which the fastest migration is allowed because the entire planetesimal disc is destabilised at once. Even in this model, the fastest possible e-folding time measusered is 
$\tau \approx 4$~Myr (Tsiganis {{\textit{et al.}}}, 2005; Morbidelli {{\textit{et al.}}}, 2005).\\

The above analysis seems to imply that the current orbital architecture of the terrestrial planets is incompatible with a late
migration of the giant planets.  However, this may not be the ultimate answer. In fact, it might be possible that the eccentricities of the terrestrial planets had been damped after the secular resonance sweeping, due to dynamical friction exerted by the planetesimals scattered by the giant planets during their migration. Moreover, as shown in Paper I, the evolution of the giant planets was not simply a smooth radial migration, as in the simulation we just presented. Potentially, the excitation of the $g_5$ mode might have happened late, relative to the $g_5=g_2$ and $g_5=g_1$ crossings. Moreover, encounters had to have happened among the gas giants and the ice giants (see Paper I), so that the radial migration of Jupiter and Saturn might not have been smooth. Also, unlike the giant planets, the terrestrial planets might not have formed on circular orbits. As we reviewed in the Introduction, the terrestrial planets formed by collisions among massive planetary embryos. As a result of collisions and encounters among massive bodies, the final orbits might have been relatively eccentric. The early simulations of this process (e.g. Chambers, 1999) predicted that the orbits of the terrestrial planets were $\sim 5$ times more eccentric than the current ones when their accretion ended. More modern simulations (e.g. O'Brien {{\textit{et al}}}., 2006), accounting for dynamical friction, succeed in producing terrestrial planets on orbits whose eccentricities and inclinations are {\it comparable} to the current values. But nothing guarantees that they had to be the {\it same} as now, as well as nothing indicates that they had to be {\it zero}. The initial orbital excitation might have been somewhat smaller than now or even larger, probably within a factor of $\sim 2$--3. This opens a new degree of freedom to be explored while addressing the effects of secular resonance sweeping. \\

Below, we will consider all these caveats, while analysing in detail each secular resonance crossing.

\section{The $g_5=g_1$ and $g_5=g_2$ resonance crossings}

In this section we discuss possibilities to alleviate or circumvent the effects of the secular resonances between the fundamental frequencies of the perihelion of Jupiter ($g_5$) and those of Venus ($g_2$) and Mercury ($g_1$). We discuss these two resonances together, because $g_2$ and $g_1$ have similar values and consequently these resonances both occur during the same phase of Jupiter's evolution. Below, we discuss in sequence four potential mechanisms: (i) the  terrestrial planets eccentricities were damped due to dynamical friction after being excited by the resonance crossing; (ii) the amplitude of the $g_5$ mode in Jupiter, $M_{5,5}$, was pumped after the secular resonances were crossed; (iii) the amplitudes of the $g_1$ and $g_2$ modes where larger originally, and they were damped {\it down} by the secular resonance crossings; (iv) Jupiter and Saturn migrated discontinuously, with jumps in semi major axes due to encounters with a Uranus-mass planet, so that the $g_5=g_1$ and $g_5=g_2$ resonances occurred very briefly or did not occur at all.

\subsection{Dynamical friction on the terrestrial planets}
\label{DF}

It might be possible that the eccentricities of the terrestrial planets evolved as in the simulation of fig.\ref{terrg5n}, but were then subsequently decreased due to dynamical friction, exerted by the large flux of planetesimals scattered by the giant planets from the outer disk. In all models (Hahn \& Malhotra, 1999; Gomes {{\textit{et al}}}., 2004; Tsiganis {{\textit{et al}}}., 2005) the mass of the planetesimal disk driving giant planet migration was 30--50$M_{\oplus}$. About a third of the planetesimals acquired orbits typical of Jupiter family comets (JFCs; perihelion distance $q<2.5$~AU) sometime during their evolution (Levison \& Duncan, 1997),
corresponding to 10--16$M_{\oplus}$. The other planetesimals remained too far from the terrestrial planets to have any influence on them. Given that the $g_5=g_2$ and $g_5=g_1$ resonances occurred when approximately 2/3 of the full migration of Jupiter and Saturn was completed (see fig.\ref{g5}), the total amount of mass of the planetesimals on JFC orbits that could exert some dynamical friction on the terrestrial planets after their excitation was about 3--5$M_{\oplus}$. \\

We investigated the magnitude of this dynamical friction in a follow-up simulation of that presented in Fig.\ref{terrg5n}. A population of 2\,000 massive objects -- with a total mass of $3.5~M_{\oplus}$ -- was added on orbits representative of the steady
state orbital distribution of JFCs (Levison \& Duncan, 1997) and the simulation was continued for 1~Myr. During that time $\sim 85\%$ of the JFCs were lost, after being scattered away by the planets. The rest survived on distant orbits. At the end of the simulation, the amplitude of the $g_2$ eccentricity term had changed only by $1.5\%$ in Earth and $4\%$ in Venus. This result is expected to scale linearly with the mass of the JFCs. Hence, to have a significant dynamical friction that can reconcile the final orbits of the terrestrial planets with the observed ones, one would need an enormous and unrealistic mass in the planetesimal population. Thus we conclude that dynamical friction cannot be the solution for the eccessive excitation of the terrestrial planets.

\subsection{Late excitation of the $g_5$ mode in Jupiter}

The resonances that are responsible for the excitation of the eccentricities of the terrestrial planets are secular resonances. Thus, their effects on the terrestrial planets are proportional to the amplitude of the $g_5$ mode in the secular evolution of the perturber i.e. Jupiter.  \\

We have seen in Paper I that the amplitude of the $g_5$ mode had to have been excited by close encounters between Saturn (or Jupiter itself) with a planet with a mass comparable to the mass of Uranus. In principle, one could think that these encounters happened relatively late, after the ratio $P_S/P_J$ exceeded 2.25 which, as shown in Fig.~\ref{g5}, corresponds to the last secular resonance crossing ($g_5=g_1$). If this were the case, when the resonance crossing occurred, the effects would have been less severe than shown in Fig.~\ref{terrg5n}. \\

In the framework of the Nice model, we tend to exclude this possibility. In the simulations performed in Tsiganis {{\textit{et al}}}. (2005) and Gomes {{\textit{et al}}}. (2005), the encounters between gas giants and ice giants start as soon as Jupiter and Saturn cross their mutual 2:1 mean motion resonance ($P_S/P_J = 2$) and end before $P_S/P_J=2.1$. In the variant of the Nice model proposed in Morbidelli {{\textit{et al}}}. (2007), the encounters start even earlier, when $P_S/P_J<2$. We do not know of any other model in which these encounters start after a substantial migration of the giant planets. \\

As a variant of this ``late $g_5$ excitation scenario'' we can also envision the possibility that the radial migration of the giant
planets and the excitation of their eccentricities started contemporarily, but the initial separation of Jupiter and Saturn was
such that $P_S/P_J \gtrsim 2.25$ from the beginning. For instance, in some of the simulations of Thommes {{\textit{et al}}}. (1999) where Uranus and Neptune are originally in between Jupiter and Saturn, initially $P_S/P_J=2.21$.  It is likely that the initial separation of Jupiter and Saturn could have been increased in these simulations to satisfy the condition $P_S/P_J \sim 2.25$, without significant changes of the results. However, the presence of Uranus and Neptune should increase the value of $g_5$ relative to that shown in Fig.~\ref{g5} for the same value of $P_S/P_J$. Hence the initial value of $P_S/P_J$ should have been even larger than 2.25 in order to avoid secular resonances with $g_1$ and $g_2$. Moreover, we have to re-iterate what we already stressed in Paper I: hydro-dynamical simulations of the evolution of the giant planets when they are embedded in the gas disk show that Jupiter and Saturn should have evolved until they got trapped into their mutual 3:2 resonance ($P_S/P_J=1.5$; Pierens \& Nelson, 2008). Initial conditions with $P_S/P_J>2.25$ are definitely inconsistent with this result. Hence, the possibility of pumping the $g_5$ mode late i.e. when $P_S/P_J \gg2.25$, should probably not be considered as a viable option.

\subsection{Decreasing the amplitudes of the $g_1$ and $g_2$ modes}
\label{down}

It is a wide-spread misconception that perihelion secular resonances excite the eccentricities (or, equivalently, the amplitudes of the resonant Fourier modes). This is true only if the initial eccentricities are close to zero: in this case, obviously, the
eccentricities can only increase. \\

The misconception comes from an un-justified use of the Lagrange-Laplace solution as an adequate integrable approximation, i.e. as the starting point for studying the full dynamics with perturbation theory. This linear approach assumes that the frequency of the eccentricity oscillations is independent of its amplitude, as is the case with a {\it harmonic oscillator}. This leads to the false
prediction that the amplitude of the resonant mode diverges to infinity at the exact resonance. In reality, however, the dynamics inside or near a secular resonance resemble those of a {\it pendulum} (see e.g.\ Chapter 8 of Morbidelli, 2002), which is a non-linear
oscillator. Strictly speaking, motion takes place {\it inside} a resonance when the corresponding resonant angle, $\phi$, {\it
librates}. Let us take as an example the motion of a test particle, with proper secular frequency $g$, perturbed by the planets. For a
resonance between $g$ and $g_k$, $\phi=(g-g_k)t+(\beta-\beta_k)$. Correlated to the librations of $\phi$ are large-amplitude {\it periodic} variations of the ``angular momentum" of the pendulum, which is a monotonically increasing function of the eccentricity of the particle. On the other hand, when we are outside but near the $g=g_k$ resonance, $\phi$ slowly circulates and the eccentricity oscillations are of smaller amplitude. Thus, each resonance has a specified {\it width}, which determines the maximum allowable excursion in eccentricity. Following this general scheme, a secular resonance between two of the planetary proper modes (e.g.\ $g_2=g_5$) can be thought of as a dynamical state, in which two modes exchange energy in a periodic fashion, according to the evolution of $\phi$. As $\phi$ moves towards one extreme of its libration cycle, one mode gains ``energy" over the other, and the eccentricity terms related to the ``winning" mode increase, in expense of terms related to the ``losing" mode. The situation for the two modes will be reversed, as $\phi$ will move towards the other extreme of the libration cycle. The total amount of ``energy'' contained in both modes has to be conserved. Hence, the eccentricity variations are determined by the initial conditions, which define a single libration curve. \\

The above scheme is correct in the conservative frame, i.e. as long as the planets do not migrate. When migration occurs, the
system may be slowly driven from a non-resonant to a resonant regime. This situation is reminiscent of the one examined in Paper I,
where the slow crossing of a mean motion resonance (MMR) was studied. However, there is a fundamental difference between the two phenomena, related to the vastly different libration time-scales. In the case of a MMR crossing, the migration rate is significantly slower than the libration frequency of the MMR. Thus, MMR crossing is an {\it adiabatic} process, and adiabatic invariance theory can be used to predict the final state of the system. When a secular resonance is crossed, things are not so simple. The crossing time is of the same order as the libration period. Thus, each moment the planets follow an ``instantaneous'' secular libration curve (i.e.\ the one that they would follow if the migration was suddenly stopped), which itself changes continuously as the planets move radially. Therefore, depending of the initial phases (i.e.,\ $\phi$) when the resonance is approached, a given eccentricity mode can decrease or increase, the final amplitude also depending on the crossing time.  If migration is very slow, the eccentricities may perform several oscillations, due to repeated libration cycles. Once the resonance is far away, the Lagrange-Laplace solution is again valid, but with different amplitudes of the former-resonant modes.  \\

From the above discussion it is clear that the result of a secular resonance crossing does not always lead to increasing the amplitude of a given secular mode. The final result depends essentially on the initial conditions (eccentricity amplitudes and phases) as well as on the migration speed. In practice, if the initial amplitude of a mode is small compared to possible eccentricity excursion along the libration curve (or, the width of the resonance) the result will be a gain in eccentricity. If, conversely, the initial eccentricity is of the order of, or larger than, the resonance width and the migration speed is not too small, then there can be a large interval of initial phases that would lead to a net loss in eccentricity. \\

With these considerations in mind, we can envision the possibility that when they formed the terrestrial planets had somewhat larger
amplitudes of the $g_1$ and $g_2$ modes than now, and that these amplitudes were damped during the $g_5$ secular resonance sweeping. 
To test this possibility, and estimate the probability that this scenario occurred, we have designed the following experiment. \\

As initial conditions for the terrestrial planets, we took the outcome of a simulation similar to that presented in Fig.~\ref{terrg5n}, so that the amplitude of the $g_2$ mode in Mercury, Venus and Earth is large ($\sim 0.1$); the initial eccentricity of Mercury is 0.12. The amplitudes of the $g_3$ and $g_4$ frequencies are, respectively, comparable ($g_3$) and smaller ($g_4$) than the current ones in all planets. Jupiter and Saturn were started with a period ratio $P_S/P_J=2.065$ and migrated to their current orbits, so that $g_5$ passes through the values of the $g_2$ and $g_1$ frequencies but avoids resonances with $g_3$ and $g_4$. As before, the migration timescale $\tau$ was assumed equal to 1~Myr and the initial eccentricities and longitudes of perihelia were chosen so that the amplitude of the $g_5$ mode in Jupiter is correct. With this set-up, we did several simulations, which differ from each other in a rotation of the terrestrial planet system relative to the Jupiter-Saturn system. This rotation changes the initial relative phase of the $g_5$ and $g_2$ terms and consequently changes the phase at which the secular resonance
$g_5=g_2$ is met. The same principle applies to the $g_5=g_1$ resonance.\\

Regardless of what happens to Mercury, we found that in about 8\% of the simulations the amplitude of the $g_2$ term in Venus was smaller than 0.025 at the end i.e. just 25\% of the initial value and comparable to the current one. \\

\begin{figure}
\resizebox{\hsize}{!}{\includegraphics[angle=-90]{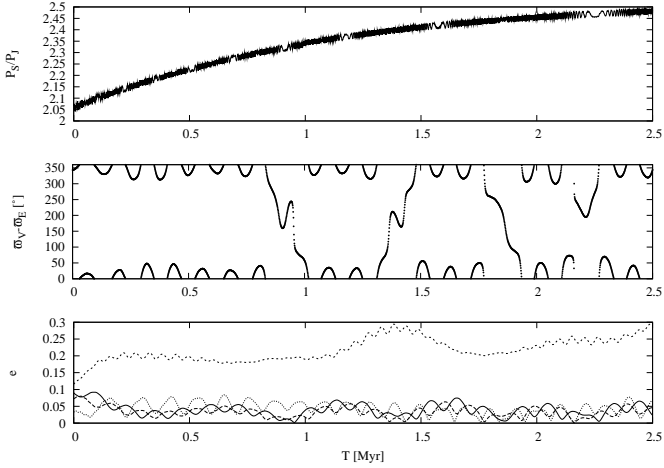}}
\caption{Evolution of the terrestrial planets during the $g_5=g_2$ and $g_5=g_1$ resonance
crossings in where the $g_2$ mode is damped down. The top panel shows
$P_S/P_J$. The middle panel shows the evolution of
$\varpi_V-\varpi_E$. The bottom panel shows the eccentricities of the
Earth (long dashed), Venus (solid), Mercury (short dash) and Mars (dotted). Notice the evident
reduction of the amplitude of oscillation of the eccentricities of
Earth and Venus near the beginning of the simulation.}
\label{g2damp} 
\end{figure}

To measure the probability that also the final orbit of Mercury is acceptable, we  used a successful simulation that damped the $g_2$
mode to construct a new series of simulations as follows.  We first measured the value of $g_5-g_2$ for the initial configuration. This was done by numerical Fourier analysis of a 8~Myr simulation with no migration imposed on Jupiter and Saturn. Defining
$P_{5,2}=2\pi/(g_5-g_2)$, we then did a simulation, still without migration of the giant planets, with outputs at multiples of
$P_{5,2}$. All these outputs had, by construction, the same relative phases of the $g_5$ and $g_2$ terms, but different relative phases of the $g_1$ term. We used five consecutive outputs, covering a full $2\pi$ range for the latter, since $P_{1,2}=2\pi/(g_2-g_1) \sim 5P_{5,2}$. Each of these outputs was used as an initial condition for a new migration simulation, with the same parameters as before. All of these simulations led essentially to the same behaviour (i.e. damping) of the $g_2$ mode, because the secular resonance $g_5=g_2$ was encountered at the same phase. But the behaviour of Mercury was different from one simulation to another. We considered the simulation successful when the eccentricity of Mercury did not exceed 0.4 during the migration simulation, as well as during a 8~Myr continuation with no migration of the giant planets; the latter was performed to determine the Fourier spectra at the end. In total we found that the evolution of Mercury satisfied these requirements in 60\% of the simulations. Figure~\ref{g2damp} shows an example of a successful simulation. In this run, the amplitude of the $g_2$ mode in Venus is damped from 0.1 to 0.025 and the amplitude of the $g_1$ mode in Mercury increases from 0.169 to 0.228.\\

In summary, we find that there is a probability of $0.08\times 0.6=4.8\%$ that the migration of the giant planets leads to a damping of the $g_2$ mode and to an acceptable final orbit of Mercury. This probability should not be taken very literally. Although it has been measured carefully, it clearly depends on the properties of the system of terrestrial planets that we start with and on the migration rate. From the considerations reported at the beginning of this section we expect that the probability decreases for reduced initial excitations of the $g_1$ and $g_2$ modes; also, the probability should decrease if slower migrations are enacted, unless the initial excitations are increased, approximately in proportion with $\tau$. We stress that the amplitude of the $g_2$ mode cannot be much larger than 0.1, otherwise the system of the terrestrial planets becomes violently unstable. Mercury is chaotic and potentially unstable on a 4~Gyr timescale even in the current system (Laskar, 1990, 1994); if the amplitude of the $g_2$ mode is larger than the current one, it becomes increasingly difficult to find solutions for Mercury that are stable for $\sim 600$~Myr, which is the putative time at which the migration of the giant planets occurred, as suggested by the LHB. \\

To conclude, we consider this scenario viable, but with a low probability to have really occurred. An additional puzzling aspect of
this mechanism that makes us sceptical, is that it requires the original amplitude of the $g_2$ mode to be much larger than that of the $g_3$ mode. As we said above, it is unclear which orbital excitation the terrestrial planets had when they formed; however, given the similarity in the masses of the Earth and Venus, nothing suggests that there should have been a significant imbalance between the amplitudes of these two modes. Actually, it would be very difficult to excite one mode without exciting the other one in a scenario in which the excitation comes form a sequence of collisions and encounters with massive planetary embryos. In fact, as explained in Paper I for the case of Jupiter and Saturn, even if only one planet has a close encounter with a third massive body, both amplitudes abruptly increase to comparable values.

\subsection{A jumping Jupiter scenario}
\label{jumping}

In Paper~I we have concluded that the current excitation of the $g_5$ and $g_6$ modes in Jupiter and Saturn could be achieved only if at least one of these planets had encounters with a body with a mass comparable to that of Uranus or Neptune. These encounters would have not just excited the eccentricity modes; they would have also provided kicks to the semi major axes of the planets involved in the encounter. In this section we evaluate the implications of this fact. \\

If Saturn scatters the ice giant onto an orbit with a larger semi-major axis, its own semi major axis has to decrease. Thus, the
$P_S/P_J$ ratio decreases instantaneously. Instead, if Saturn scatters the ice giant onto an orbit with smaller semi major axis, passing it to the control of Jupiter, and then Jupiter scatters the ice giant onto an orbit with larger semi major axis, the semi major axis of Saturn has to increase, that of Jupiter has to decrease, and consequently $P_S/P_J$ has to increase. This opens the possibility that $P_S/P_J$ jumps, or at least evolves very quickly, from less than 2.1 to larger than 2.25, thus avoiding
secular resonances between $g_5$ and $g_2$ or $g_1$ (see Fig.~\ref{g5}).\\

We now turn to the Nice model, because this is our favoured model and the one on which we have data to do more quantitative analysis. 
In the Nice model, only in a minority of the successful runs (i.e. the runs in which both Uranus and Neptune reach stable orbits at locations close to the current ones) there are encounters between Jupiter and an ice giant without ejecting it. For instance, this happened in one simulation out of six in Gomes {{\textit{et al}}}. (2005), and in two simulations out of 14 in Nesvorn\'{y} {{\textit{et al}}}. (2007). So, the probability seems to be of the order of 15\%. In all other runs, only Saturn encounters an ice giant, which, as we said above, decreases the $P_S/P_J$ ratio instead of increasing it. \\

We point to the attention of the reader that Nesvorn\'{y} {{\textit{et al}}}. (2007) showed that encounters with Jupiter would explain the capture of the irregular satellites of this planet and their orbital properties. If Jupiter never had encounters, only Saturn, Uranus and Neptune should have captured irregular satellites (unless Jupiter captured them by another mechanism, but then it would be odd that the system of the irregular satellites of Jupiter is so similar to those of the other giant planets; see Jewitt \& Sheppard, 2006). Moreover, in the Nice model, the cases where Jupiter has encounters with an ice giant are those which give final values of $P_S/P_J$ which best approximate the current Solar System, whereas in the other cases Saturn tends to end its evolution a bit too close to the Sun (Tsiganis {{\textit{et al}}}., 2005). These two facts argue that, although improbable, this kind of dynamical evolution, involving Jupiter encounters with an ice giant, actually occurred in the real Solar System. \\

\begin{figure}
\resizebox{\hsize}{!}{\includegraphics[angle=-90]{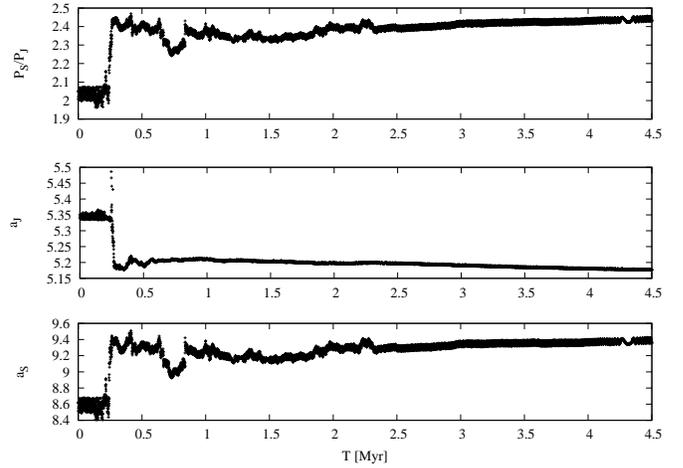}}
\caption{Evolution of $P_S/P_J$ (top), $a_J$ (middle) and $a_S$
  (bottom) in a Nice-model simulation in which Jupiter has close
  encounters with Uranus.} 
\label{pl3} 
\end{figure}

As an example of what can happen in the Nice model when Jupiter encounters and ice giant, the top panel of Fig.~\ref{pl3} shows the
evolution of $P_S/P_J$; the middle panel displays the evolution of the semi-major axis of Jupiter and in the bottom panel that of Saturn. The time resolution of the output is 100~yr, though only every 1\,000~years is shown here. This new simulation is a ``clone'' of one of the original simulations of Gomes {{\textit{et al.}}}  (2005). The positions and velocities of the planets were taken at a time when Jupiter and Saturn had just crossed their 2:1 MMR, but had not yet experienced encounters with the ice giants -- this time is denoted by $t=0$ in Fig.~\ref{pl3} and corresponds to $t=875.5$~Myr in the simulation of Gomes {{\textit{et al.}}} (2005) where the giant planet instability occured at a time that roughly corresponds to the chronology of the LHB. \\

\begin{figure}
\resizebox{\hsize}{!}{\includegraphics[angle=-90]{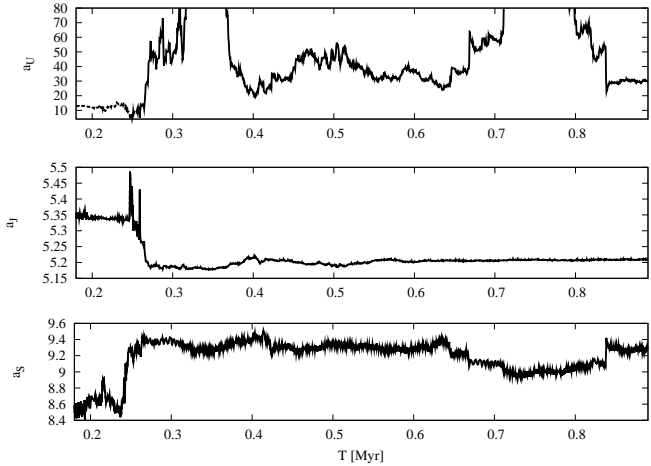}}
\caption{Magnification of the dynamics of Fig.~\ref{pl3}, in the
  180,000y--890,000y interval. The top panel now shows the evolution
  of Uranus' semi major axis.}
\label{pl3mag} 
\end{figure}

As one can see in Fig~\ref{pl3}, $P_S/P_J$ evolves very rapidly from $P_S/P_J<2.1$ to $P_S/P_J>2.4$ around $t= 0.25$~Myr. Then $P_S/P_J$ decreases again below 2.4, has a rapid incursion into the 2.3-2.4 interval around $t= 0.75$~Myr and eventually, after $t=2.5$~Myr, increases smoothly to 2.45. The latter value is a very good approximation of the current value of $P_S/P_J$. The striking similarity of the curves in the top and middle panels demonstrates that the orbital period ratio is essentially dictated by the dynamical evolution of Saturn. Nevertheless, Saturn can have this kind of early evolution only if Jupiter has encounters with an ice giant, which justifies the name of ``jumping Jupiter scenario'' used in this section. This can be understood by looking at the evolution of the two planets and of Uranus in details, as we describe below with the help of the magnification of the dynamics, which is provided in Fig.~\ref{pl3mag}. \\

Saturn first has an excursion in semi-major axis from approximately 8.5~AU to 8.8~AU at $t= 0.215$~Myr. This happens because it has repeated encounters with Uranus, which lead to a temporary exchange of their orbits, placing Uranus at $a=8.4$~AU. Then Saturn kicks Uranus back outwards, which again puts Saturn at $\sim 8.5$~AU. This series of events shows that, if Jupiter does not participate in this phase of encounters, the dynamics are characterized by energy exchange between Saturn and Uranus: if one planet is scattered outwards, the other is scattered inwards and vice-versa. Given that Uranus was initially much closer to the Sun than it is now, the net effect on Uranus had to be an outward scattering; consequently, in the absense of encounters with Jupiter, Saturn would have had to move inwards. So, this evolution could not have led to an increase in $P_S/P_J$. \\

The situation is drastically different if encounters between Jupiter and Uranus occur. In Fig.\ 5, this starts to happen at $t\sim 0.25$~Myr), when Saturn's semi-major axis evolves rapidly to $\sim 9.2~$AU and Uranus' semi-major axis to $\sim 6.5~$AU. Jupiter first exchanges orbits with Uranus: Jupiter moves out to 5.52~AU while $a_U$ reaches 3.65~AU while the perihelion distance of Uranus, $q_U$, decreases to $\sim 2~$AU. However, notice that the intrusion of Uranus into the asteroid belt is not a necessary feature of the jumping Jupiter scenario; some simulations leading only to $q_U \sim 3$--$4$ AU. Then Jupiter scatters Uranus outwards to $a_U\sim 50$~AU, itself reaching $a_J\sim 5.2$~AU.  The situation is now very different from the one before: Uranus is back on a trans-Saturnian orbit and, because this was the result of a Jupiter-Uranus encounter, Saturn has not moved back to its original position. Thus, this series of encounters has lead to an irreversible increase of the orbital separation (and period ratio) of Jupiter and Saturn. The subsequent evolution of the planets, shown in Fig.~\ref{pl3mag}, is dominated by encounters between Saturn and Uranus. These encounters push Uranus' semi-major axis to beyond 200 AU at $t= 0.35$~Myr and to beyond 100 AU at $t=$ 0.725--0.775~Myr, but in both cases Saturn pulls it back. This erratic motion of $a_U$ correlates with the one of $a_S$. Eventually Uranus' semi-major axis stabilises at $\sim 35$~AU. Thus, Uranus and Neptune switch positions, relative to their initial configuration. This happened in all our
simulations where Jupiter-Uranus encounters took place. \\

We now proceed to simulate the evolution of the terrestrial planets in the framework of the evolution of Jupiter, Saturn and Uranus discussed above. However, we cannot simply add the terrestrial planets in the system and redo the simulation because the dynamics is
chaotic and the outcome for the giant planets would be completely different. Thus, we need to adopt the strategy introduced by Petit
{\it et al.} (2001). More precisely, we have modified the code Swift-WHM (Levison \& Duncan,1994), so that the positions of Jupiter,
Saturn and Uranus are computed by interpolation from the output of the original simulation. Remember that the orbital elements of the outer planets had been output every 100~yr. The interpolation is done in orbital element space, and the positions and velocities are computed from the result of the interpolation. For the orbital elements $a, e, i, \Omega$, and $\omega$, which vary slowly, the interpolation is done linearly. For the mean anomaly $M$, which cycles over several periods in the 100~yr output-interval, we first
compute the mean orbital frequency from the mean semi major axis (defined as the average between the values of $a$ at the beginning and at the end of the output-interval) and then adjust it so that $M$ matches the value recorded at the end of the output-interval. Then, over the output-interval, we propagate $M$ from one time-step to another, using this adjusted mean orbital frequency.\\

To test the performance of this code, we have done two simulations of the evolution of the 8 planets of the solar system. In the first one, the planets were started from their current configuration and were integrated for 1~Myr, using the original Swift-WHM code. In the second one, an encounters phase was simulated, by placing Uranus initially in between the orbits of Jupiter and
Saturn, while setting the terrestrial planets on circular and co-planar orbits, and integrating for 1~Myr.  In both simulations the orbital elements of the planets were recorded every 100~yr. Then, we re-integrated the terrestrial planets, in both configurations, using our new code, in which the orbital evolutions of Jupiter, Saturn and Uranus were being read from the output of the previous integration. We then compared the outputs of the two simulations, for each initial planet configuration. Eccentricity differences between the two simulations of the same configuration are interpreted here as the ``error" of our new code. In the first configuration (current system), which represents a quite regular evolution, we found that the root-mean-square errors in eccentricity are $1.05\times 10^{-4}$ for Mercury, $4.3\times 10^{-5}$ for Venus, $4.0\times 10^{-5}$ for the Earth and $8.9\times 10^{-5}$ for Mars. In the second simulation (repeated encounters) the evolution of the terrestrial planets, Mercury in particular, is more violently chaotic. Consequently, the error remains acceptable ($\sim 3\times 10^{-3}$) for all planets except Mercury, for which the error grows above $\sim 5\times 10^{-3}$ after $\approx 0.5~$Myr. This is due to a shift in the secular phases of Mercury's orbit in the two simulations, which changes the outcome of the evolution. Given the chaotic character of the dynamics, both evolutions are equaly likely and acceptable. Thus, we conclude that our modified integrator is accurate enough (although the effects of encounters among the giant planets are ``smeared'' over 100~yr intervals) to be used effectively for our purposes. \\

\begin{figure}
\resizebox{\hsize}{!}{\includegraphics[angle=-90]{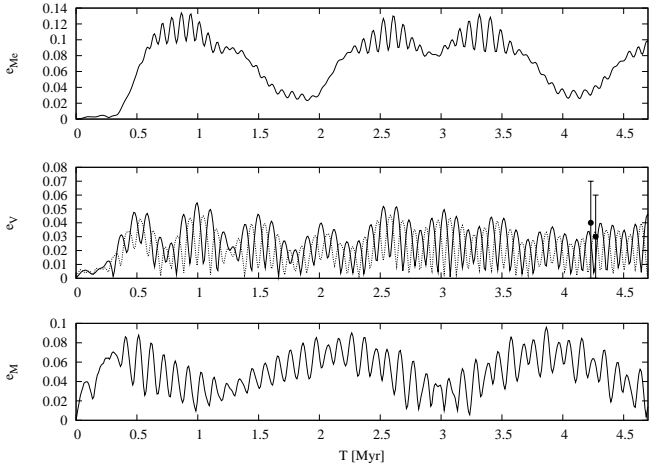}}
\caption{The evolution of the eccentricities of Mercury (top), Venus
  and Earth (respectively solid and dashed lines) in the middle panel, and
  Mars (bottom), during the dynamics of Jupiter and Saturn illustrated
  in Fig.~\ref{pl3}. The initial orbits of the terrestrial planets are
  assumed to be circular and coplanar. The solid circles and vertical bars represent 
the eccentricity oscillation of Earth and Venus, as in Fig.~\ref{terrg5n}.}
\label{pl3circTP} 
\end{figure}

For a comparison with the case illustrated in Fig.~\ref{terrg5n}, we first present a simulation where all terrestrial planets start from coplanar, circular orbits. Fig.~\ref{pl3circTP} shows the evolution of the eccentricity of Mercury (top), Venus and Earth (middle) and Mars (bottom). The eccentricities of the terrestrial planets increase rapidly but, unlike in the case of smooth migration of Jupiter and Saturn (Fig.~\ref{terrg5n}), they remain moderate and do not exceed the values characterising their current secular evolutions (see for instance Laskar, 1990). A Fourier analysis of a 4~Myr continuation of the simulation, with Jupiter and Saturn freely evolving from their final state, gives amplitudes of the $g_2$ and $g_3$ modes in Venus and the Earth of $\sim 0.015$, in good agreement with the real values. The amplitude of the $g_4$ mode in Mars is smaller than the real one. For Mercury, the analysis is not very significant because the $g_1$ and $g_5$ frequencies are closer to each other than in reality (because $P_S/P_J$ is a bit smaller, which makes $g_5$ faster, and the inclination of Mercury is a bit larger, which makes $g_1$ smaller). Nevertheless, in a 20~Myr simulation, the eccentricity of Mercury does not exceed 0.4.\\

\begin{figure}
\resizebox{\hsize}{!}{\includegraphics[angle=-90]{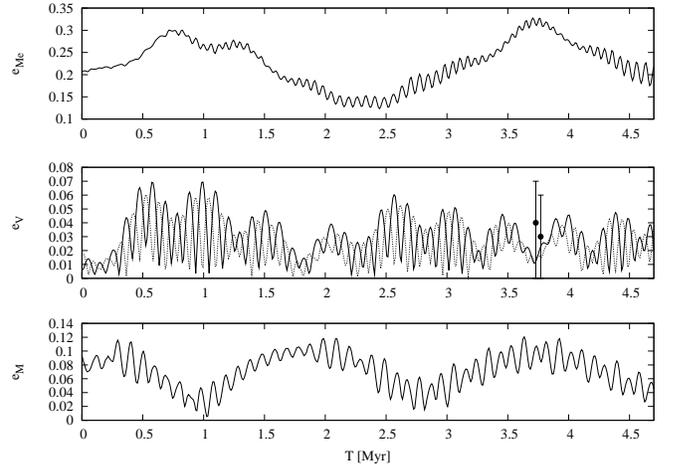}}
\caption{The same as Fig.~\ref{pl3circTP}, but for terrestrial planets
  starting from their current orbits.}
\label{pl3currentTP} 
\end{figure}

Since it is unlikely that the terrestrial planets formed on circular orbits, but from the beginning should have had some orbital
excitation, remnant of their violent formation process, we have re-enacted the evolution of the terrestrial planets, but starting from their current orbits. In this case we assume that their current orbital excitation is an approximation of their primordial excitation. Again, Jupiter and Saturn evolve as in Fig.~\ref{pl3}. The results are illustrated in Fig.~\ref{pl3currentTP}. We  find that the eccentricities of the terrestrial planets remain moderate, and comparable to the current values. A Fourier analysis of the continuation of this simulation shows again that amplitudes of the $g_2$ and $g_3$ terms in Venus and the Earth are $\sim 0.015$. The final amplitude of the $g_4$ term in Mars has preserved the initial value of  $\sim 0.04$. The maximal eccentricity of Mercury does not exceed 0.35. \\

Taken together, these two simulations are a successful demonstration that the rapid evolution of  $P_S/P_J$  over the 2.1-2.4 range
allows the excitation of the terrestrial planets to remain small, because the sweeping of the $g_5$ secular resonance is too fast to have a noticeable effect. \\ 

We have to stress, though, that not all ``jumping-Jupiter'' evolutions are favourable for the terrestrial planets. In some cases the rapid evolution of $P_S/P_J$ ends when the orbital period ratio is $\sim 2.2$ or less, which is close to the $g_5=g_1$ or $g_5=g_2$ resonance. In other cases $P_S/P_J$, after having increased to above 2.3, decreases again and remains for a long time in the range of values for which these secular resonances occur. In these cases, the destiny of the terrestrial planets is set: Mercury typically becomes unstable, and the Earth and Venus become much more eccentric than they are in reality, due to the excitation of the $g_2$ mode. \\

\begin{figure}
\resizebox{\hsize}{!}{\includegraphics[angle=-90]{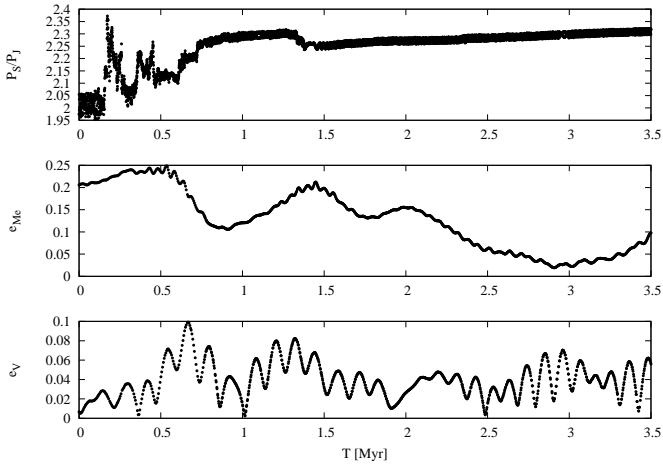}}
\caption{Top panel: the evolution of the period ratio between Saturn
  and Jupiter in another ``jumping Jupiter'' evolution from the Nice
  model. Middle and bottom panel: the evolutions of the eccentricities
  of Mercury and Venus, respectively. }
\label{pl0currentTP} 
\end{figure}

Nevertheless, the evolution of $P_S/P_J$ does not need to be as fast as in Fig.~\ref{pl3} to lead to ``good'' terrestrial planets. Fig.~\ref{pl0currentTP} gives another example from a different realisation of the Nice model: the top panel shows the evolution of $P_S/P_J$, the middle and the bottom panel show the evolutions of the eccentricities of Mercury and Venus, respectively, for a system of terrestrial planets starting from current orbits. In this case $P_S/P_J$ evolves rapidly to $\sim 2.35$, but then it returns in the 2.10-2.25 range, where it spends a good half-a-million years. Subsequently, after evolving again to 2.3, it returns to 2.25 and starts to increase again, slowly. In this case the effect of secular resonances is no longer negligible as in the case illustrated above. But in our terrestrial planet simulation the eccentricity of Mercury is effectively damped, enacting the principle formulated in sect.~\ref{down}. The eccentricity of Venus, via the amplitude of the $g_2$ mode, is excited during the time when $P_S/P_J\sim 2.15$ and reaches 0.1. However, it is damped back when $P_S/P_J$ decreases again to 2.5 at $t\sim 1.5$~Myr. A the end, the orbits of the terrestrial planets are again comparable to their observed orbits, in terms of eccentricity excitation and amplitude of oscillation. \\

At this point, one might wonder what is the fraction of giant planets evolutions in the Nice model that are favourable for the
terrestrial planets. This is difficult to evaluate, because we did only a limited number of simulations and then cloned the simulations that seemed to be the most promising. We try, nevertheless, to give a rough estimate. As we said at the beginning of this section, the jumping Jupiter evolutions are about 15--20\% of the successful Nice model runs. By successful we mean those runs that at the end yielded giant planets on orbits resembling their observed ones, without considering terrestrial planets constraints. The successful runs are about 50--70\% of the total runs (Gomes {\it et al.}, 2005; Tsiganis {{\textit{et al}}}., 2005). Most of the unsuccessful runs were of the jumping Jupiter category, but led to the ejection of Uranus. It is possible that, if the planetesimal disc had been more massive than the one used in the simulation, or represented by a larger number of smaller particles, so as to better resolve the process of dynamical friction, Uranus would have been saved more often, thus leading to a larger fraction of successful jumping Jupiter cases. Then, by cloning jumping Jupiter simulations after the time of the first encounter with Uranus, we find that about 1/3 of the ``successful'' jumping Jupiter cases are also successful for the terrestrial planets, like the cases presented in Figs.\ 6, 7 and 8; $P_S/P_J$ evolves very quickly beyond 2.3, and does not exceed 2.5 in the end. This brings the total probability of having in the Nice model a giant planet evolution compatible with the current orbits of the terrestrial planets to about 6\%.
 
\section{The $g_5=g_3$ and $g_5=g_4$ resonance crossings}

The study described in this section is very similar to that presented at two DPS conferences by Agnor (2005) and Agnor \& Lin (2007) and therefore its results are not new. However, Agnor and collaborators never presented their work in a formal publication so, for 
completeness, we do it here. \\

As discussed in Section~2, when Jupiter and Saturn are close enough to their mutual 2:1 mean motion resonance, the $g_5$ frequency can be of the order of 17--18 $\arcsec$/yr (see Fig.~\ref{g5}), and therefore resonances with the proper frequencies of Mars and the Earth ($g_4$ and $g_3$) have to occur. If the migration of Jupiter and Saturn from the 2:1 resonance is sufficiently fast, as shown in
Fig.~\ref{terrg5n}, the sweeping of the $g_5=g_4$ and $g_5=g_3$ does not cause an excessive excitation of the amplitudes of the $g_4$ and $g_3$ mode. \\

However, while a fast departure from the 2:1 resonance is likely, in the original version of the Nice model (Gomes {{\textit{et al}}}., 2005) Jupiter and Saturn have to {\it approach} the 2:1 resonance very slowly, in order to cross the resonance with sufficient
delay to explain the timing of the LHB. During the approach phase the eccentricities of Jupiter and Saturn (and, therefore, also the amplitude of the $g_5$ term) were small but, nevertheless, the long timescales involved could allow the secular resonances to have a
destabilising effect on the terrestrial planets. \\

\begin{figure}
\resizebox{\hsize}{!}{\includegraphics[angle=-90]{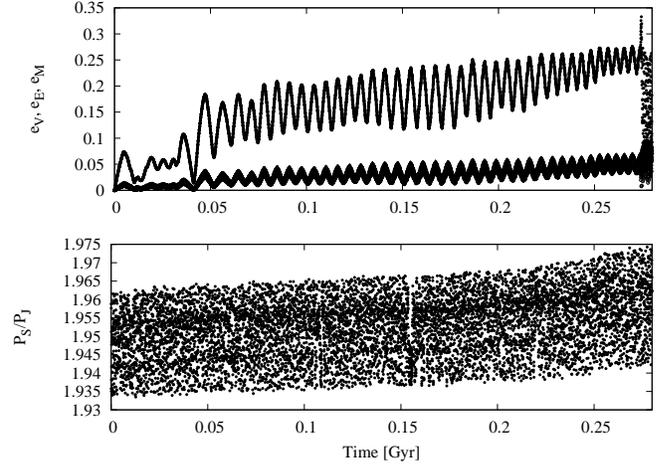}}
\caption{The evolution of eccentricities of Venus, Earth and Mars 
  in a simulation where Jupiter and Saturn approach slowly
  their 2:1 resonance. The top panel shows the evolution of the
  eccentricities of Mars (upper curve) and of Earth and Venus
  (overlapping almost perfectly in a unique lower curve). The bottom
  panel shows $P_S/P_J$ vs. time.}
\label{marsunstable} 
\end{figure}

To illustrate this point, we repeated the last part of the simulation presented in Gomes {{\textit{et al}}}. (2005), during which Jupiter and Saturn slowly approach the 2:1 resonance, and we added the terrestrial planets, Venus to Mars, initially on circular orbits. The top panel of Fig.~\ref{marsunstable} shows the evolution of the eccentricities of the terrestrial planets, while on the bottom panel the evolution of the period ratio between Saturn and Jupiter over the integrated time-span is depicted. The initial configuration is very close to the $g_5=g_4$ resonance. At this time, Jupiter's eccentricity is oscillating from $\sim 0$ to $\sim 0.015$, but with a fast frequency related to the 2:1 resonance with Saturn. The amplitude of the $g_5$ mode in Jupiter is only $3\times 10^{-4}$. Eventually, the secular resonance (i.e. at $t\sim 0.045$~Gyr) excites the amplitude of the $g_4$ term in Mars to approximately 0.15, in Earth and Venus to 0.033 and 0.023 respectively, as well as the amplitude of the $g_5$ mode in Jupiter to $2\times 10^{-3}$. At the same time, the $g_4$ frequency increases abruptly because of Mars' larger eccentricity to become approximately 0.2$\arcsec$/y faster than $g_5$. The long periodic oscillations of the eccentricity of Mars after 0.05 Gyr are precisely related to the $g_4-g_5$ beat. As migration proceeds, the $g_5$ frequency increases. Surprisingly, the $g_4$ frequency increases accordingly, which causes the mean eccentricities of Mars, Earth and Venus to increase accordingly. We interpret this behavior as if the dynamical evolution is sticking to the outer separatrix of the $g_5=g_4$ resonance, in apparent violation of adiabatic theory. Therefore the $g_5=g_4$ resonance is not crossed again. At the end, the eccentricity of Mars becomes large enough to drive the terrestrial planets into a global instability. This is a very serious problem, which seems to invalidate the original version of the Nice model. \\

Fortunately, the new version of the Nice model, presented in Morbidelli {{\textit{et al}}}. (2007) solves this problem. This version was built to remove the arbitrary character of the initial conditions of the giant planets that characterised the original version of the model. In Morbidelli {{\textit{et al}}}. (2007), the initial conditions of the N-body simulations are taken from the output of an hydrodynamical simulation in which the four giant planets, evolving in the gaseous proto-planetary disk, eventually reach a non-migrating, fully resonant, stable configuration. More precisely, Jupiter and Saturn are trapped in their mutual 3:2 resonance (see also Masset \& Snellgrove, 2001; Morbidelli \& Crida, 2007; Pierens \& Nelson, 2008); Uranus is in the 3:2 resonance with Saturn and Neptune is caught in the 4:3 resonance with Uranus. Morbidelli {{\textit{et al}}}. (2007) showed that, from this configuration, the evolution is similar to that of the original version of the Nice model (e.g.\ Tsiganis {{\textit{et al}}}., 2005; Gomes {{\textit{et al}}}., 2005), but the instability is triggered when Jupiter and Saturn cross their mutual 5:3 resonance (instead of the
2:1 resonance as in the original version of the model). Additional simulations done by our group (Levison {{\textit{et al}}}., in preparation) show that, if the planetesimal disk is assumed as in Gomes {{\textit{et al}}}. (2005), the instability of the giant planets is triggered as soon as a pair of planets leaves their original resonance, and this can happen as late as the LHB chronology seems to suggest. \\

For our purposes in this paper, the crucial difference between the new and the original version of the Nice model is that Jupiter and Saturn approach their mutual 2:1 resonance fast, because the instability has been triggered before, when Jupiter and Saturn were still close (or more likely locked in) their 3:2 resonance. \\

To test what this implies for the terrestrial planets, we have run several simulations, in which Jupiter and Saturn, initially on
quasi-circular orbits, migrate all the way from within the 5:3 resonance to the 2:1 resonance on timescales of a several million
years (corresponding to $\tau=5$--25~Myr). We stress that we enact a smooth migration of the giant planets (evolutions of the jumping Jupiter case would be a priori more favourable) and that the migration time to the 2:1 resonance could easily be much shorter than we assume (see for instance the bottom panel of Fig. 8 in Morbidelli {{\textit{et al}}}., 2007). On the other hand, in these simulations the amplitude of the $g_5$ term in Jupiter is excited through the multiple mean motion resonance crossings between Jupiter and Saturn, and therefore is about 1/3 of the current value (see Paper~I). \\

\begin{figure}
\resizebox{\hsize}{!}{\includegraphics[angle=-90]{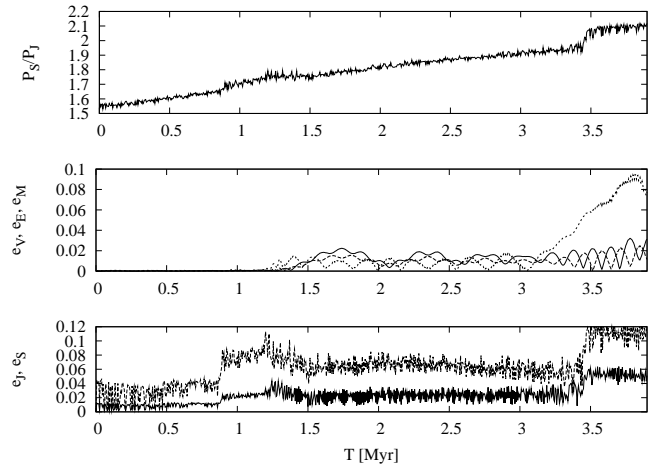}}
\caption{The evolution of eccentricities of Venus, Earth and Mars,
  starting from circular orbits,  
  in a simulation where Jupiter and Saturn migrate from within the 5:3
  resonance to the 2:1 resonance. The top panel shows ratio of orbital
  periods between Saturn and Jupiter. The middle panel shows the
  evolution of the eccentricities of Mars (short dashed), Earth
(long dashed) and Venus (solid). The bottom panel
  shows the eccentricities of Jupiter (solid) and Saturn (dashed).}
\label{nice21} 
\end{figure}

\begin{figure}
\resizebox{\hsize}{!}{\includegraphics[angle=-90]{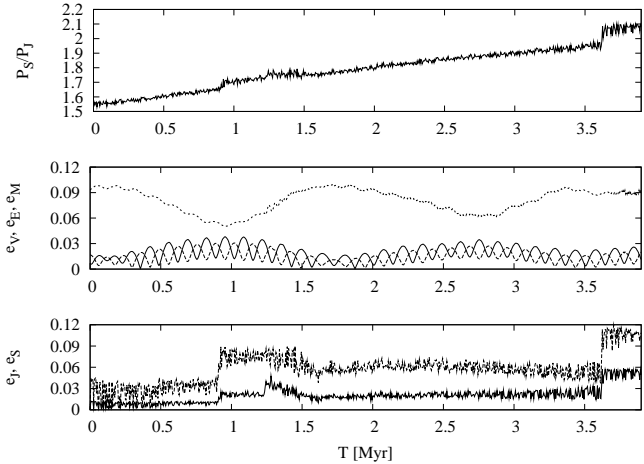}}
\caption{The same as Fig.~\ref{nice21}, but for terrestrial planets 
starting from their current orbits. Notice that the evolution for
Jupiter and Saturn is not exactly the same as that of the previous
simulation, despite the initial conditions and $\tau$ are the
same. This is a result of the effects of resonance crossings, which
are chaotic and thus give effects that are different, from run to run.}
\label{nice22} 
\end{figure}

We present two simulations here. Both have been obtained with $\tau=5$~Myr; the first one investigates the outcome of initially circular terrestrial planets while the other uses the current orbits of the terrestrial planets. The results are presented in Figures~\ref{nice21} and~\ref{nice22}, respectively. In both figures, the top panel shows the evolution of the orbital period ratio between Saturn and Jupiter; the middle panel shows the eccentricities of the terrestrial planets except Mercury, and the bottom panel the eccentricities of Jupiter and Saturn. \\

Notice the little jumps of the $P_S/P_J$ curve and of $e_J$ and $e_S$ when Jupiter and Saturn cross mean motion resonances. Also notice that in the interim between resonances, the eccentricities of Jupiter and Saturn decay, as a consequence of a damping term that we introduced on Saturn to mimic the effect of dynamical friction, and to ensure that they do not become unstable (see Paper~I). In the case where the terrestrial planets have initially circular orbits (Fig.~\ref{nice21}) we see that they start to be excited after 1.5 Myr, when Jupiter and Saturn cross their mutual 9:5 resonance and become more eccentric themselves. The simultaneous increase in the eccentricities of Venus and Earth is mostly caused by an increase in their $g_2$ terms because of the near-resonance $g_5=g_2$ when $P_S/P_J \sim 1.7$ (see fig.~\ref{g5}); the amplitude of their $g_2$ terms becomes approximately 0.02. The effect of the $g_5=g_4$ resonance is visible towards the end of the simulation, when Mars becomes suddenly more eccentric (dashed line; middle panel) and has its $g_4$ term increased to about 0.08. Its eccentricity does not exceed 0.09, however, and decreases again in response of the giant planets crossing the 2:1 resonance. In the next figure, the case where the terrestrial planets are initially on their current orbits (fig.~\ref{nice22}), we see almost no changes in the eccentricity evolution of the terrestrial planets, which remain of the same order as the initial ones. The most likely reason for this behaviour is that the chaotic nature of the migration causes the behaviour of Jupiter and Saturn to be slightly different from one simulation to the next. Indeed, the value of $P_S/P_J$ has a longer plateau in fig.~\ref{nice21} than in fig.~\ref{nice22}. In addition, in the first simulation the value of $M_{5,5}$ is slighly larger than in the second, which can be seen from the slightly larger excursions of Jupiter's eccentricity from its mean. In addition, the little changes in the eccentricities of the terrestrial planets in fig.~\ref{nice21} become nearly invisible if the eccentricities are already quite large.\\

Thus, we conclude that the $g_5=g_4$ and $g_5=g_3$ secular resonances are not a hazard for the terrestrial planets, at least in the new version of the Nice model. 

\section{Note on the inclinations of the terrestrial planets}

The dynamical excitation of the terrestrial planets at the end of the process of formation is still not known. Simulations (e.g. O'Brien {{\textit{et al}}}., 2006) show that the planets had an orbital excitation comparable to the current one, but this could be an artifact of the poor modeling of dynamical friction. Therefore, any indirect indication of what had to be the real dynamical state of the orbits of the terrestrial planets would be welcome. \\

In the two previous sections we have seen that original circular orbits cannot be excluded. In fact, if the terrestrial planets had
circular orbits after their formation, the current eccentricities and amplitudes of the secular modes could have been acquired during the evolution of the giant planets through secular resonance crossings and reactions to the jumps in eccentricity of Jupiter's orbit. It is interesting to see if the same is true for the inclinations, which are between 2 and 10 degrees for the real terrestrial planets. In fact, if the terrestrial planets had originally quasi-circular orbits (probably as a result of strong dynamical friction with the planetesimals remaining in the inner solar system at the end of the giant collisions phase), presumably they had also inclinations close to zero. If the evolution of the giant planets could not excite the inclinations of the terrestrial planets, then  the orbits of these planets had to be excited from the very beginning. In the opposite case, co-planar (and circular) initial orbits cannot be excluded. \\

Unlike the eccentricity case, there are no first-order secular resonances affecting the evolution of the inclinations of the
terrestrial planets during the migration of Jupiter and Saturn. In fact, neglecting Uranus and Neptune, the secular motion of Jupiter and Saturn in the Lagrange--Laplace theory is characterised by a unique frequency ($s_6$). Its value is now approximately $-26\arcsec$/yr and should have been faster in the past, when the two planets were closer to each other. The frequencies of the longitudes of the node of the terrestrial planets are all smaller in absolute value than $s_6$: the frequencies $s_1$ and $s_2$ are about $-6$ to $-7\arcsec$/yr; the $s_3$ and $s_4$ frequencies are $-18$ to $-19\arcsec$/yr (Laskar, 1990). So, no resonances of the kind $s_6=s_k$, with $k=1$ to 4 were possible. \\

\begin{figure}
\resizebox{\hsize}{!}{\includegraphics[angle=-90]{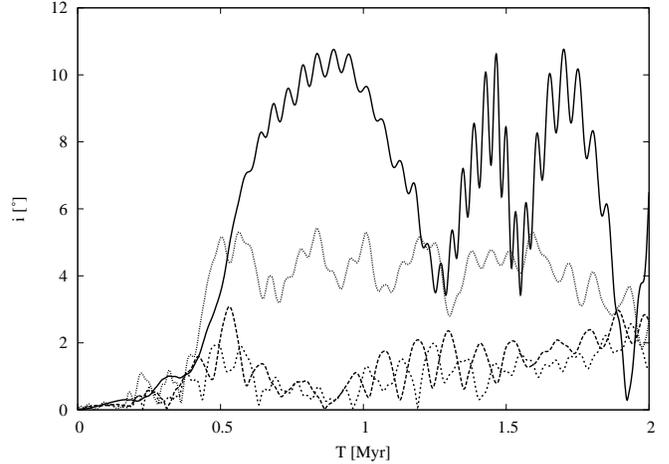}}
\caption{Evolution of the inclinations of Mercury (solid),
    Venus (dark dashed) Earth (light dashed) and Mars (dotted), all starting from 0, in the
    ``jumping Jupiter'' evolution shown in the top panel of
    Fig.~\ref{pl0currentTP}.}
\label{incls} 
\end{figure}

Nevertheless, in our simulations with terrestrial planets starting with circular and co-planar orbits in section~\ref{jumping} we find
(see Fig.~\ref{incls}) that the inclinations are excited and the current values can be reproduced, including that of the Mercury (whose mean real inclination is $\sim 8\degr$; Laskar, 1988) and Mars (4$\deg$). This is because there are a number of non-linear secular resonances that can occur, which also involve Uranus or Neptune. The frequencies $s_1$ and $s_2$ could resonate with $s_7$ or $s_8$, when/if the latter were larger in absolute value than the current values (respectively $-2.99$ and $-0.67\arcsec$/yr now), which is likely when the ice giants had much stronger interactions with Jupiter and Saturn. The frequencies $s_3$ and $s_4$ were close to the 1:2 resonance with $s_6$, when the latter was larger than now due to the smaller orbital spacing between the giant planets. As is well known, a resonance $s_6 = 2 s_4$ cannot have dynamical effects, because the corresponding combination of the angles does not fulfil the D'Alembert rule (see for instance Morbidelli, 2002).  However, a resonance like $2 s_4=s_6+s_7$ does satisfy the D'Alembert rule and is close to the previous one, due to the current small value of $s_7$. The variety of dynamical evolutions of Uranus and Neptune that are possible in the Nice model (even within the subset of jumping-Jupiter evolutions) precludes us to say deterministically which resonances really occurred, when and with which effects. But the possibility of exciting the current
inclinations starting from $i\sim 0$ is not remote and therefore, unfortunately, we cannot conclude on what had to be the initial dynamical state of the orbits of the terrestrial planets.
 
\section{Discussion and Conclusions}

In this work we have shown that the radial smooth migration of Jupiter and Saturn tends to excite the eccentricities of the orbits of Mercury, Venus and the Earth well above the values achievable during their current evolution. This happens because the $g_5$ frequency decreases during the migration; consequently it enters temporarily in resonance with $g_2$ and $g_1$ when $P_S/P_J\sim 2.1$--2.3, which excites the amplitudes associated with these two frequencies in the Fourier spectrum of the terrestrial planets. Conversely, the amplitude of the $g_4$ frequency in Mars is not excited too much, provided that Jupiter and Saturn approached and departed rapidly to/from their mutual 2:1 mean motion resonance, as in the new version of the Nice model (Morbidelli {{\textit{et al}}}., 2007). \\

We have found two possible, but low-probability mechanisms that may make giant planet migration compatible with the current orbital
architecture of the terrestrial planets. One requires that the original structure of the terrestrial planets was quite strange,
with an amplitude of the $g_2$ mode significantly larger than that of the $g_3$ mode. In this case, the $g_5=g_2$ resonance could have
damped the amplitude of the $g_2$ mode (and to some extent also of the $g_1$ mode), for some lucky combination of secular phases. The other requires that $P_S/P_J$ ``jumped'' (or evolved very rapidly across) the 2.1--2.3 interval, as a result of encounters between {\it both} Jupiter and Saturn with either Uranus or Neptune. Some evolutions of this kind occur in the Nice model, but they are rare (successful probability approximately $5\%$). \\

We are aware that most readers will consider this a first serious drawback of the Nice model. But, before leaving way to critics, we would like to advocate some relevant points. \\

First, this apparent problem is not confined to the Nice model, but concerns any model which associates the origin of the LHB to a delayed migration of Jupiter and Saturn (e.g. Levison {{\textit{et al}}}., 2001, 2004; Strom {{\textit{et al}}}., 2005; Thommes {{\textit{et al}}}., 2007).\\

Second, an easy way out of this problem is to deny that the LHB occurred as an impact spike. In this case, giant planet migration
might have occurred as soon as the gas disk disappeared, without consequences on the terrestrial planets, which had not formed yet. We warn against this superficial position. It seems to us that there are at least four strong pieces of evidence in favour of the cataclysmic spike of the LHB: (i) basins on the Moon as big as Imbrium and Orientale could not have formed as late as 3.8~Gyr ago if the bombardment rate had been declining monotonically since the time of planet formation at the rate indicated by dynamical models without giant planet migration (Bottke {{\textit{et al}}}., 2007); (ii) old zircons on Earth demonstrate that the climate on Earth 4.3--3.9~Gyr ago was relatively cool (i.e. the impact rate was low; Mojzsis {{\textit{et al}}}., 2001), and that strong heating events, probably associated with impacts, happened approximately 3.8~Gyr ago (Trail {{\textit{et al}}}., 2007); (iii) the most prominent impact basins on Mars occurred after the disappearance of Martian magnetic field (Lillis {\it et al.}\ 2006, 2007); (iv) impact basins on Iapetus occurred after the formation of its equatorial ridge, which is estimated to have formed between $t=$200-800~Myr (Castillo-Rogez {{\textit{et al}}}., 2007). Anybody seriously arguing against the cataclysmic nature of the LHB should find an explanation for each of these issues. \\

Third, one may argue that the origin of the LHB was not determined by a delayed migration of the giant planets, but that it was caused
instead by dynamical events that concerned only the inner solar system. The model by Chambers (2007) does precisely this and, in our
opinion, it is the most serious alternative to the Nice model, for the origin of the LHB. In Chambers' scenario, the system of terrestrial planets originally contained five planets. The fifth rogue planet, of sub-martian mass, was in between the current orbit of Mars and the inner edge of the asteroid belt. The orbit of this planet became unstable at a late epoch: after crossing the asteroid belt for some time and dislodging most of the asteroids originally resident in that region (which caused the LHB) the rogue planet was eventually removed by a collision with another planet, a collision with the Sun, or ejection from the solar system. Although appealing, this model has never been given much consideration and has never been tested in detail, for instance against the current structure of the asteroid belt and the magnitude of the LHB. We stress that Chambers' model cannot by any means explain the bombardment of Iapetus. Thus, to accept this model one has to find an alternative explanation of the late bombardent of this satellite of Saturn that does not involve or imply a migration of the giant planets or, alternatively, to prove that its ridge formed much earlier than estimated by Castillo-Rogez {{\textit{et al}}}. (2007), so that the heavy bombardment of Iapetus could have occurred early. \\

As a final note, we remark that it is very dangerous to exclude or adopt a model based on probabilistic arguments on events that concern the habitability of the Earth, such as its orbital excitation. The fact that we are here to study these problems introduces an obvious observational bias: of all possible solar systems, we can be only in one that allows our existence, however improbable was the chain of events that led to its formation. The Dolar System is full of low-probability properties related to habitability: the presence of the Moon (necessary to stabilise the obliquity of the Earth; Laskar {{\textit{et al}}}., 1993), the quasi-circular orbits of the giant planets (as opposed to the extra-solar planets), the absence of giant planets in the temperate zone, to quote only a few. Therefore we think that the low probability to preserve a moderately excited orbit of the Earth cannot be used to disqualify the Nice model.

\begin{acknowledgements}
This work is part of the Helmholtz Alliance's 'Planetary evolution and Life', which RB and AM thank for financial support. Exchanges
between Nice and Thessaloniki have been funded by a PICS programme of France's CNRS, and RB thanks the host KT for his hospitality during a recent visit. RG thanks Brasil's CNPq and FAPERJ for support. HFL is thankful for support from NASA's OSS and OPR programmes. Most of the simulations in this work were performed on the CRIMSON Beowulf cluster at OCA.
\end{acknowledgements}

\section{References}
Agnor, C., Canup, R., Levison, H. 1999 Icarus 142, 219. \\
Agnor, C. 2005 DPS 37, 29.01 \\
Agnor, C. \& Lin, D. BAAS 38, 537.\\
All\`egre, C., Manh\`{e}s, G. \& G\"{o}pel, C. E\&PSL 267, 386. \\
Bottke, W., Levison, H., Nesvorn\'{y}, D. \& Dones, L. 2007 Icarus 190, 203. \\
Castillo-Rogez, J.~C., Matson, D.~L., Sotin, C., Johnson, T.~V., Lunine, J.~I., \& Thomas, P.~C. 2007 Icarus 190, 179.\\
Chambers, J. \& Wetherill, G. 1998 Icarus 136, 34. \\
Chambers, J. 1999 MNRAS 304, 793. \\
Chambers, J. 2001 Icarus 152, 205. \\
Chambers, J. 2007 Icarus 189, 386. \\
Correia, A. \& Laskar, J. 2004 Nature 429, 848.\\
Gomes, R., Morbidelli, A. \& Levison, H. 2004 Icarus 170, 492.\\
Gomes, R., Levison, H., Tsiganis, K. \& Morbidelli, A. 2005 Nature 435, 466.\\
Hahn, J. \& Malhotra, R. 1999 AJ 117,3041\\
Haisch K.E., Lada E.A. \& Lada C.J. 2001 ApJ 553, 153. \\ 
Hartmann, W. K., Ryder, G., Dones, L., \& Grinspoon, D. 2000 In: 'Origin of the Earth and Moon`` (R. Canup \& R. Knighter, eds). Univ. Arizona Press, Tucson, Arizona.  \\
Jewitt, D. \& Sheppard, S. 2005 Sp. Sci. Rev. 116, 441. \\
Kenyon, S. \& Bromley, B. 2006 AJ 131, 1837. \\
Kne\v{z}evi\'{c}, Z., Milani, A., Farinella, P, Froeschle, Ch. \& 
Froeschle, Cl. 1991, Icarus 93, 316. \\
Laskar, J. 1988 A\&A 198, 341. \\
Laskar, J. 1990 Icarus 88, 266. \\
Laskar, J., Joutel, F. \& Robutel, P. 1993 Nature 361, 615. \\
Laskar, J. 1994 A\&A 287, L9. \\
Laskar, J. 2008 Icarus 196, 1.\\
Levison, H. \& Duncan, M. 1994 Icarus 108, 18 \\
Levison, H.~F.\& Duncan, M.~J.\ 1997.\  Icarus 127, 13.\\ 
Levison, H. F., Dones, L., Chapman, C. R., Stern, S. A., Duncan, M. J. \& Zahnle, K. 2001 Icarus 151, 286 \\
Levison, H., Morbidelli, A. \& Dones, L. 2004 AJ 128, 2553. \\
Lillis, R.~J., Frey, H.~V., Manga, M., Mitchell, D.~L., Lin, R.~P., Acu{\~n}a, M.~H. 2007 LPI 
Contributions 1353, 3090. \\
Lillis, R.~J., Frey, H.~V., Manga, M., Halekas, J.~S., Mitchell, D.~L., Lin, R.~P.\ 2006 AGU Fall 
Meeting Abstracts A5. \\
Malhotra, R. 1995 AJ 110, 420\\
Masset F. \& Snellgrove, M. 2001 MNRAS 320, 55\\
Minton, D. \& Malhotra, R. 2009 Nature 457, 1109. \\
Mojzsis, S.~J., Harrison, T.~M., \& Pidgeon, R.~T. 2001 Nature 409, 178. \\
Morbidelli, A. 2002 Modern Celestial Mechanics - Aspects of Solar System Dynamics (Taylor \& Francis, UK). \\
Morbidelli A. \& Crida, A. 2007 Icarus 191, 158\\
Morbidelli, A., Tsiganis, K., Crida, A., Levison, H. \& Gomes, R. 2007 AJ 134,1790 \\
Murray, C. \& Dermott, S. 1999 Solar System Dynamics (Cambridge University Press, Cambridge, UK) \\
Nesvorn\'{y}, D., Vokrouhlick\'{y}, D. \& Morbidelli, A. 2007 AJ 133, 1962 \\
Nobili, A. \& Will, C. Nature 320, 39. \\
O'Brien, P., Morbidelli, A. \& Levison, H.. 2006 Icarus 184, 39.\\
Petit, J.-M., Morbidelli, A. \& Chambers, J. 2001. Icarus 153, 338.\\ 
Pierens, A. \& Nelson, R. 2008 A\&A 482, 333\\
Raymond, S.~N., Quinn, T., \& Lunine, J.~I. 2004 Icarus 168, 1. \\
Raymond, S.~N., Quinn, T., \& Lunine, J.~I. 2005 ApJ 632, 670. \\
Raymond, S.~N., Quinn, T., \& Lunine, J.~I. 2006 Icarus 183, 265.\\
Raymond, S.~N., Quinn, T., \& Lunine, J.~I. 2007 Astrobiology 7, 66. \\
Ryder, G., Koeberl, C \& Mojzsis, S. 2000. In: 'Origin of the Earth and Moon`` (R. Canup \& R. Knighter, eds). Univ. Arizona Press, Tucson, Arizona. \\
Saha, P. \& Tremaine, S. 1994 AJ 108, 1962. \\
Strom, R. G., Malhotra, R., Ito, T., Yoshida, F., \& Kring, D. A. 2005 Science 309, 1847.\\
Tsiganis, K., Gomes, R., Morbidelli, A. \& Levison, H.  2005 Nature 435, 459\\
Thommes, E., Duncan, M. \& Levison, H. 1999 Nature 402, 635\\
Thommes, E., Nilsson, L \& Murray, N. 2007 ApJ 656, 25\\
Touboul, M., Kleine, T., Bourdon, B., Palme, H., \& Wieler, R. 2007 Nature 450, 1206\\
Trail, D., Mojzsis, S., Harrison, T. 2007 GeCoA 71, 4044. \\
Wieczorek, M.~A., Le Feuvre, M., Rambaux, N., Laskar, J., \& Correia, A.~C.~M. 2009 LPISCA 40, 1276
\end{document}